\documentclass[lettersize,journal]{IEEEtran}
\usepackage{amsmath,amsfonts}
\usepackage{algorithmic}
\usepackage{algorithm}
\usepackage{array}
\usepackage[caption=false,font=normalsize,labelfont=sf,textfont=sf]{subfig}
\usepackage{textcomp}
\usepackage{stfloats}
\usepackage{url}
\usepackage{verbatim}
\usepackage{graphicx}
\usepackage{cite}
\usepackage{soul}  
\usepackage{xcolor} 
\begin{document}

\title{12-bit Delta-Sigma ADC operating at a temperature \\ of up to 250~°C in Standard 0.18~$\mu$m SOI CMOS}

\author{Christian Sbrana,~\IEEEmembership{Student Member,~IEEE,} Alessandro Catania,~\IEEEmembership{Senior Member,~IEEE,} Tommaso Toschi, Sebastiano Strangio,~\IEEEmembership{Senior Member,~IEEE,} Giuseppe Iannaccone,~\IEEEmembership{Fellow,~IEEE}
\thanks{
{\bf This work has been submitted to the IEEE for possible publication. Copyright may be transferred without notice, after which this version may no longer be accessible.}

This research was partially supported by the CHARM project from the ECSEL Joint Undertaking (Grant Agreement N.876362), by the European Union’s Horizon 2020 research and innovation programme and by the Italian Ministry for Economic Development (MISE), and by the ECS4DRES project supported by the Chips Joint Undertaking (grant agreement number 101139790) and its members, including the top-up funding by Germany, Italy, Slovakia, Spain and The Netherlands. It is also partially supported by the italian MUR under the Forelab project of the ``Dipartimenti di Eccellenza'' programme.

Christian Sbrana is with Quantavis s.r.l., Largo Padre Renzo Spadoni, 56126 Pisa, Italy and with the Dipartimento di Ingegneria dell'Informazione (DII), Università di Pisa, 56122 Pisa, Italy;
Alessandro Catania is with the Dipartimento di Ingegneria dell'Informazione (DII), Università di Pisa, 56122 Pisa, Italy;
Tommaso Toschi is with Quantavis s.r.l., Largo Padre Renzo Spadoni, 56126 Pisa, Italy;
Sebastiano Strangio is with the Dipartimento di Ingegneria dell'Informazione (DII), Università di Pisa, 56122 Pisa, Italy;
Giuseppe Iannaccone is with the Dipartimento di Ingegneria dell'Informazione (DII), Università di Pisa and with Quantavis s.r.l., Largo Padre Renzo Spadoni, 56126 Pisa, Italy (corresponding author:
(e-mail: giuseppe.iannaccone@unipi.it)}
\thanks{Manuscript submitted December, 2024.}}

{

\IEEEpubid{0000--0000/00\$00.00~\copyright~2021 IEEE}

\maketitle

\begin{abstract}
Some applications require electronic systems to operate at extremely high temperature. Extending the operating temperature range of automotive-grade CMOS processes --- through the use of dedicated design techniques --- can provide an important cost-effective advantage.
We present a second-order discrete-time delta-sigma analog-to-digital converter operating at a temperature of up to 250~°C, well beyond the 175~°C qualification temperature of the automotive-grade CMOS process used for its fabrication (XFAB XT018). The analog-to-digital converter incorporates  design techniques that are effective in mitigating the adverse effects of the high temperature, such as increased leakage currents and electromigration. We use configurations of dummy transistors for leakage compensation, clock-boosting methods to limit pass-gate cross-talk, and we optimized the circuit architecture to ensure stability and accuracy at high temperature.
Comprehensive measurements demonstrate that the analog-to-digital converter achieves a signal-to-noise ratio exceeding 93 dB at 250~°C, with an effective number of bits of 12, and a power consumption of only 44~mW. The die area of the converter is only 0.065~mm$^2$ and the area overhead of the high-temperature mitigation circuits is only 13.7\%. The Schreier Figure of Merit is 140~dB at the maximum temperature of 250~°C, proving the potential of the circuit for reliable operation in challenging applications such as gas and oil extraction and aeronautics.
\end{abstract}



\section{Introduction}

\IEEEPARstart{T}{he} push for the digitization of industrial systems, aimed at improving control and overall system performance, often requires the use of electronics in harsh environments. 
Several industrial applications, such as aerospace, aeronautics, oil and gas drilling, energy generation, motor control, and some types of manufacturing, have much higher operating temperature than conventional electronics, \cite{Neudeck2002,Hassan2018,Werner2001} as illustrated in Figure \ref{fig_HTapps}. 
Indeed, electronic components and CMOS circuits for consumer and most industrial applications are qualified for operation at a maximum temperature of 85-125~°C, whereas those for military and automotive applications are qualified at up to 150-175~°C.

\IEEEpubidadjcol

The challenge of high-temperature operation can be addressed in three different ways: 
\begin{description}
    \item[i)] by segregating electronics in cooler areas or by proper thermal insulation of the electronic subsystem, paying a price in terms of longer wiring, larger weight and volume, and higher cooling costs \cite{Cecchi2024}; 
    \item[ii)] by using devices and circuits based on wide bandgap materials such as SiC and GaN, that are inherently suitable for high-temperature operation up to 600~°C \cite{Hassan2018, Iannaccone2021}, and using high-temperature materials also for wires, bonding, packaging and printed circuit board\cite{Shaddock2015, Pradhan2024};
    \item[iii)] by extending the temperature range of automotive-grade silicon-on-insulator (SOI) CMOS processes up to 250-300~°C, through the use of design techniques that compensate for the degradation mechanisms occurring at high temperature \cite{Bugakova2024,Sbrana2024}; SOI-CMOS has reduced substrate leakage currents due to the insulating layer between the substrate and the active area.
\end{description}
\begin{figure}[b!]
\centerline{\includegraphics[width=0.5\textwidth]{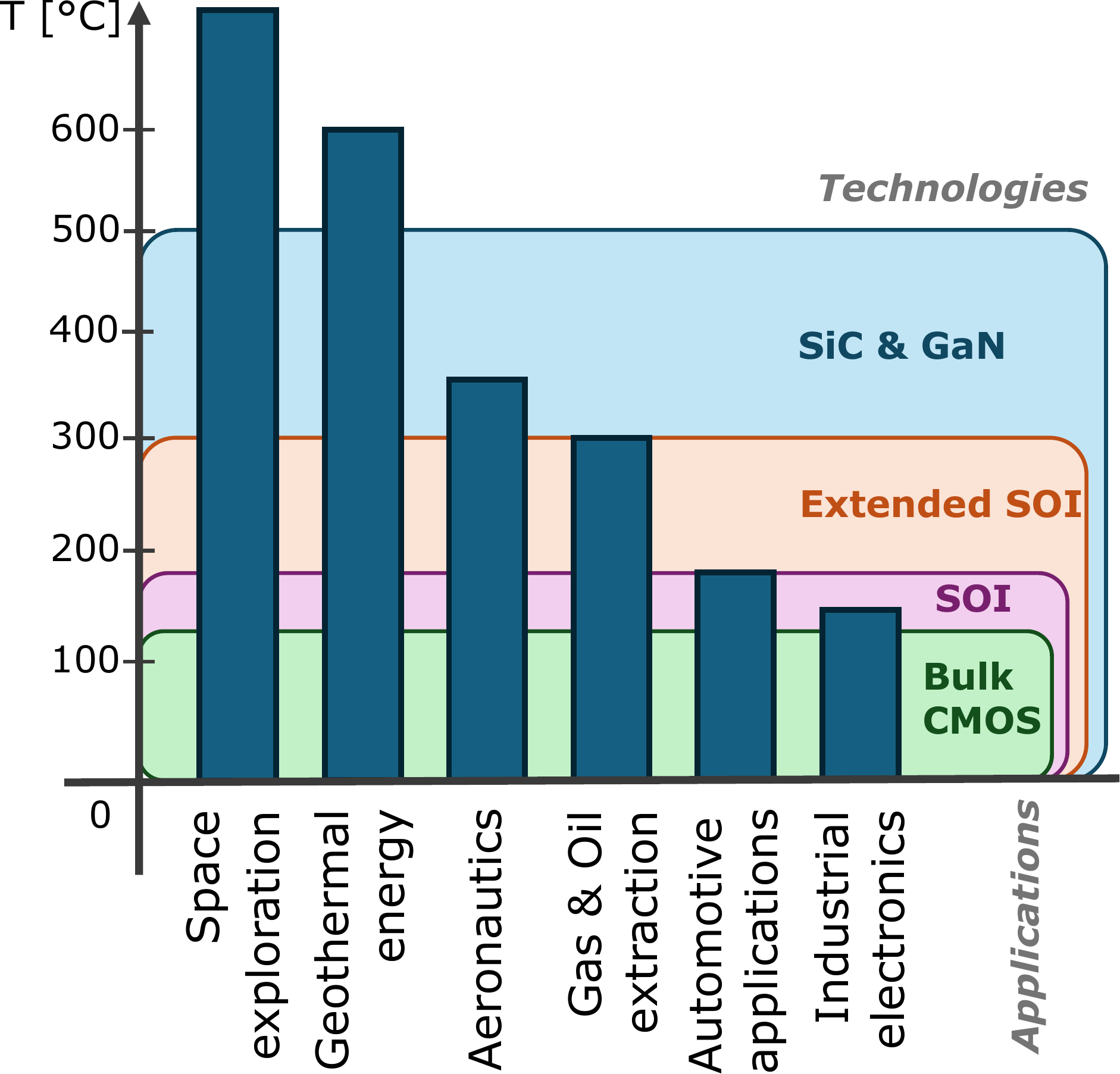}}
\caption{Available ICs manufacturing technologies match the temperature ratings of different applications. Data extracted from Ref. \cite{Neudeck2002,Shaddock2015,Werner2001}.
}
\label{fig_HTapps}
\end{figure}
The first option may suffer from reliability issues caused by the associated deployment complications and may not be suitable in some industrial settings where form factors are prohibitive.
The second option provides a promising and definitive solution to the issue, when reliable and reasonably scaled-down SiC or GaN monolithic processes will be available \cite{Buffolo2024} (results have only been presented with SiC CMOS processes with minimum feature size of 1~$\mu$m \cite{Young2013,Romijn2022,Kuhns2016,Rahman2016}).
The third option has the advantage of being based on available and cost-effective monolithic fabrication processes, and is suitable for many industrial applications (Figure \ref{fig_HTapps}). 

Basic proof of concepts have been provided up to 300~°C for single transistors \cite{Karulkar1993,Grella2013}, focusing also on the analysis of the effects induced by the high temperature on MOSFET parameters \cite{Rudenko2002,Flandre1993}.
Basic digital and analog circuits have been shown and their potential for high-temperature operations have been investigated \cite{Demeus1998,Demeus1999,Grella2013,Francis1992,Kappert2015}, sometimes providing a useful comparison with other existing technologies \cite{Demeus2001} or giving important design guidelines for robust circuits realization \cite{Eggermont1996}.
SOI CMOS technology has been identified as the solution of choice for integrated microsystems designed to operate in high-temperature environments with very different requirements of power consumption and speed \cite{Flandre2001}.

In this paper, we present a second-order discrete-time (DT) delta-sigma ($\Delta$-$\Sigma$) analog-to-digital converter (ADC) designed for high-temperature operation up to 250°C. 
It is realized using an automotive-grade CMOS process, the XFAB XT018~\cite{XFAB}, a 180 nm partially depleted SOI CMOS technology with aluminum metal interconnections. 
The process is qualified in the range from -40~°C to 175~°C, thus the process design kit (PDK) models are calibrated in the same temperature range, likely leading to a poor device behavior prediction above 175~°C. For this reason, a high-temperature-specific design approach is needed to enable proper operation above this limit.

In \cite{Sbrana2024}, using some basic test structures, we have already investigated the possibility of exploiting the same SOI platform above the maximum temperature rating, by characterizing the degradation of device electrical parameters and proposing dedicated design countermeasures. The presented ADC is based on the requirements of a high-accuracy conversion of a relatively slow input signal, since the ADC is part of a larger IC for application in temperature measurements for the reliable monitoring of thermal processes in industrial ovens.  


In the technical literature, $\Delta$-$\Sigma$~ADCs have been demonstrated up to 175~°C in a 0.18~$\mu$m SOI CMOS \cite{Korotkov2020} and up to 250°C in a 0.35~$\mu$m High Temperature SOI CMOS process \cite{Kappert2022}, however at the expense of large area occupation and power consumption. Also SAR converter architectures perform well up to 210~°C at high sampling frequency \cite{Watson2014,AD7981}, but the performance degradation is still significant when approaching 300°C \cite{Zou2015}. 
With respect to results from the literature, our proposed ADC has a extremely low area occupation and energy consumption, for an ENOB in excess of 12 bits up to 250~°C, and a very good Schreier Figure of Merit (FoM) of 140 dB at 250~°C\cite{Schreier2005} defined as 
${\rm FoM} \equiv {\rm SINAD} + 10 \times \log (BW/P)$,
where SINAD is the signal-to-noise-and-distortion ratio, $BW$ is the bandwidth and $P$ is the power consumption

We discuss the high-temperature ADC architecture and the design techniques used to mitigate the effects of leakage currents in Section II. Then we describe the measurement setup (Section III) and the electrical characterization of the converter over the whole temperature range (Section IV), comparing our results with the state of the art. We draw the Conclusion in Section V.


\section{$\Delta$-$\Sigma$ ADC temperature-aware design}

\subsection{Converter architecture}

Due to critical design issues emerging at high temperature \cite{Sbrana2024}, it is recommended to focus on as simple as possible architectures.
Considering the specifications, a second-order DT $\Delta$-$\Sigma$ ADC has been identified as the best solution. 
In fact, in contrast to a first-order modulator architecture, the oversampling ratio (OSR) of a second-order modulator can be tuned to optimize the speed-resolution trade-off. 
The system utilizes a single-bit cascaded integrator feedback modulator, which offers a favourable trade-off among hardware complexity, power consumption, and resolution, without imposing strict requirements on the matching characteristics of the digital-to-analog converter (DAC) \cite{Schreier2005}.

Although the  $\Delta$-$\Sigma$ architecture has been largely adopted for very challenging applications with rigid design constraints on supply voltage \cite{Catania2022}, speed \cite{Li2019} and resistance to radiation \cite{Edwards1999}, we want to stress that this architecture is not inherently designed for high-temperature applications, and thus ad-hoc design optimization becomes necessary.

Fig. \ref{fig_ADCmod}(a) shows the simplified circuit-level diagram of the proposed second-order $\Delta$-$\Sigma$ modulator. 
A digital low-pass filter (not shown) has been implemented on the chip to obtain a full $\Delta$-$\Sigma$ ADC.
The two-phase driving clock method is employed to control both the first and the second integrators with precise timing, enabling the capacitors to effectively sample and convert the input signals. 
The result is a single-bit bitstream signal (BTS), produced by the comparator based on the differential feedback signal which can assume either V\textsubscript{DD} or GND levels.

The integrator stages use a fully differential switched-capacitor topology, based on large gain–bandwidth product (GBW) folded-cascode amplifiers.
A low-power dynamic common-mode feedback (CMFB) circuit is exploited to stabilize the output common-mode voltage of the folded-cascode amplifiers.

In a two-stage filter design, the first stage must focus on minimizing input-referred noise and offset. 
Provided that the thermal ($kT/C$) noise is a critical concern in switched-capacitor topologies, priority has been assigned to the first integrator by allocating larger capacitors, amplifier area, and power consumption (capacitors in the first stage have a capacitance that is ten times larger than that of capacitors in the second stage)

\begin{figure*}[!t]
\centering
\centerline{\includegraphics[width=2.0\columnwidth]{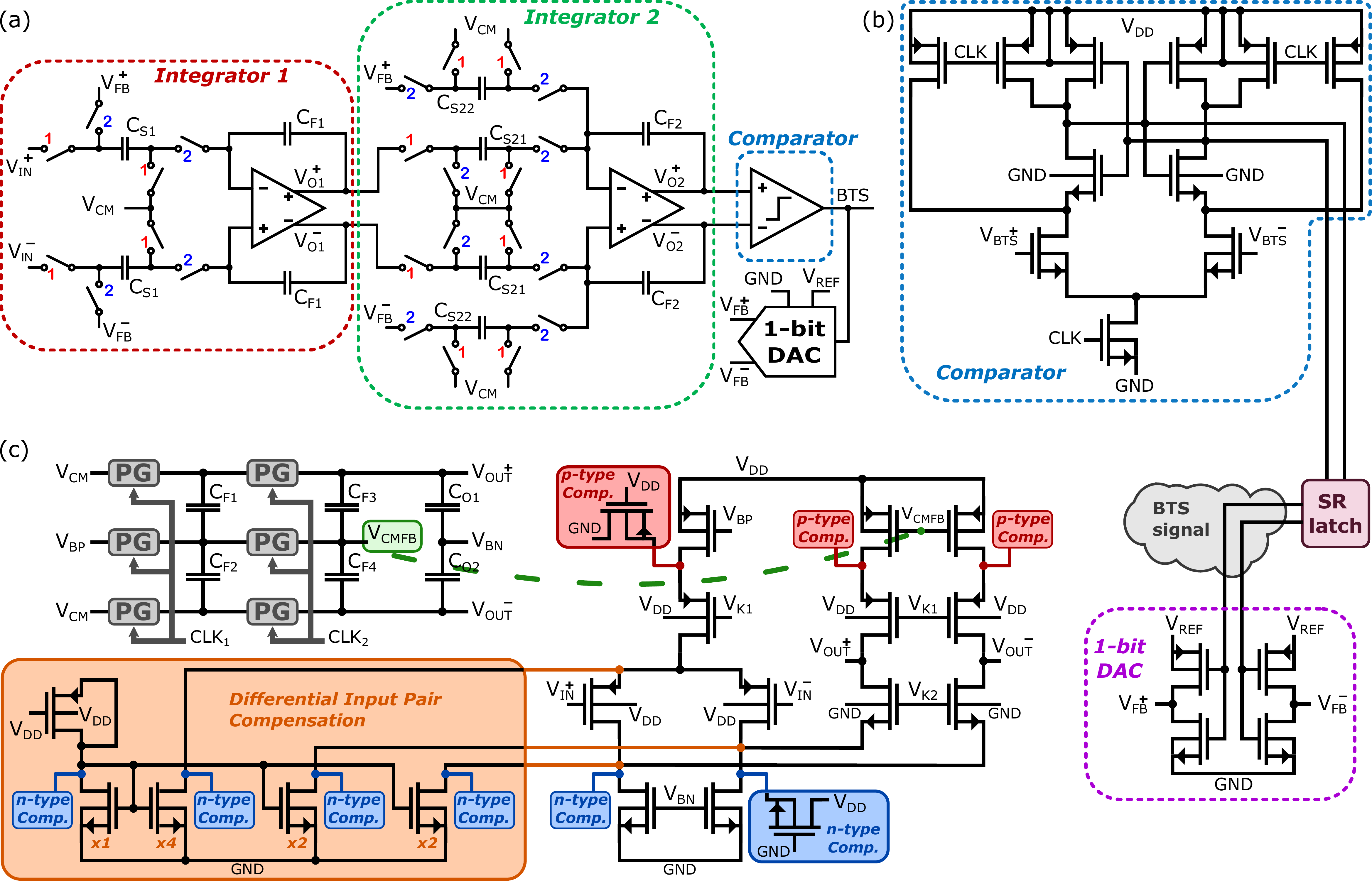}}
\caption{ADC architecture schematic overview: (a) 2\textsuperscript{nd}-order DT Delta-Sigma modulator; (b) StrongArm latch comparator for bitstream signal output and single-bit DAC for feedback; (c) Folded-cascode operational amplifier used in the integrators, with junction leakage compensation method for single PMOS (p-type) and NMOS (n-type), and for the differential input pair. The dynamic CMFB circuit is shown in the same subpicture, providing the bias voltage V\textsubscript{CMFB} for the cascode amplifier, based on leakage-compensated pass-gates (PG).}
\label{fig_ADCmod}
\end{figure*}

Finally, a StrongARM latch architecture has been selected for the comparator \cite{Razavi2015}, reported in Fig. \ref{fig_ADCmod}(b), offering the benefits of negligible static power consumption and no hysteresis.
At the output, an SR latch provides a robust digital output to drive the DAC with a single-bit topology.


\subsection{Design techniques for high-temperature operation}


The main high-temperature challenges in the integrated circuit design are the increased leakage currents in the silicon regions and the electromigration in automotive-grade metal lines \cite{Sbrana2024}.
The leakage currents with the most significant impact are the reverse currents of the p-n junctions and the subthreshold channel leakage.
Electromigration affects the metal lines, with a progressive increase of the interconnect resistance until open circuit, due to microscopic damage in conditions of high current density and high temperature.


The reverse saturation current of p-n junctions, from now on called junction leakage current, has a near-exponential increase with temperature, leading to higher power consumption and potentially triggering latch-up phenomena \cite{Sbrana2024}. 
The negative leakage effect introduced by drain (and source) to body junctions can be mitigated using dummy-transistor compensation \cite{Wang2021}. 
This strategy is based on the idea of neutralizing the leakage currents of both the source-body and drain-body junctions by injecting an identical but opposite current.
This extra contribution is obtained by adding parallel off-state dummy transistors, with the body and source connected and enforcing a reverse bias on the drain-body junction.

Fig. \ref{fig_ADCmod}(c) shows this junction leakage compensation method applied to the folded-cascode amplifier. 
A n-type dummy transistor is used to compensate the leakage current of a NMOS with source-body connection to the ground, realizing the current cancellation by exploiting its drain-body current, when the gate is grounded and the supply voltage $V\textsubscript{DD}$ is applied to the drain.
Similarly, a p-type dummy transistor is used for PMOS compensation with source-body connection to the supply voltage, now applying to the dummy gate $V\textsubscript{DD}$ and GND to the drain. 

\begin{figure}[t!]
\centerline{\includegraphics[width=\columnwidth]{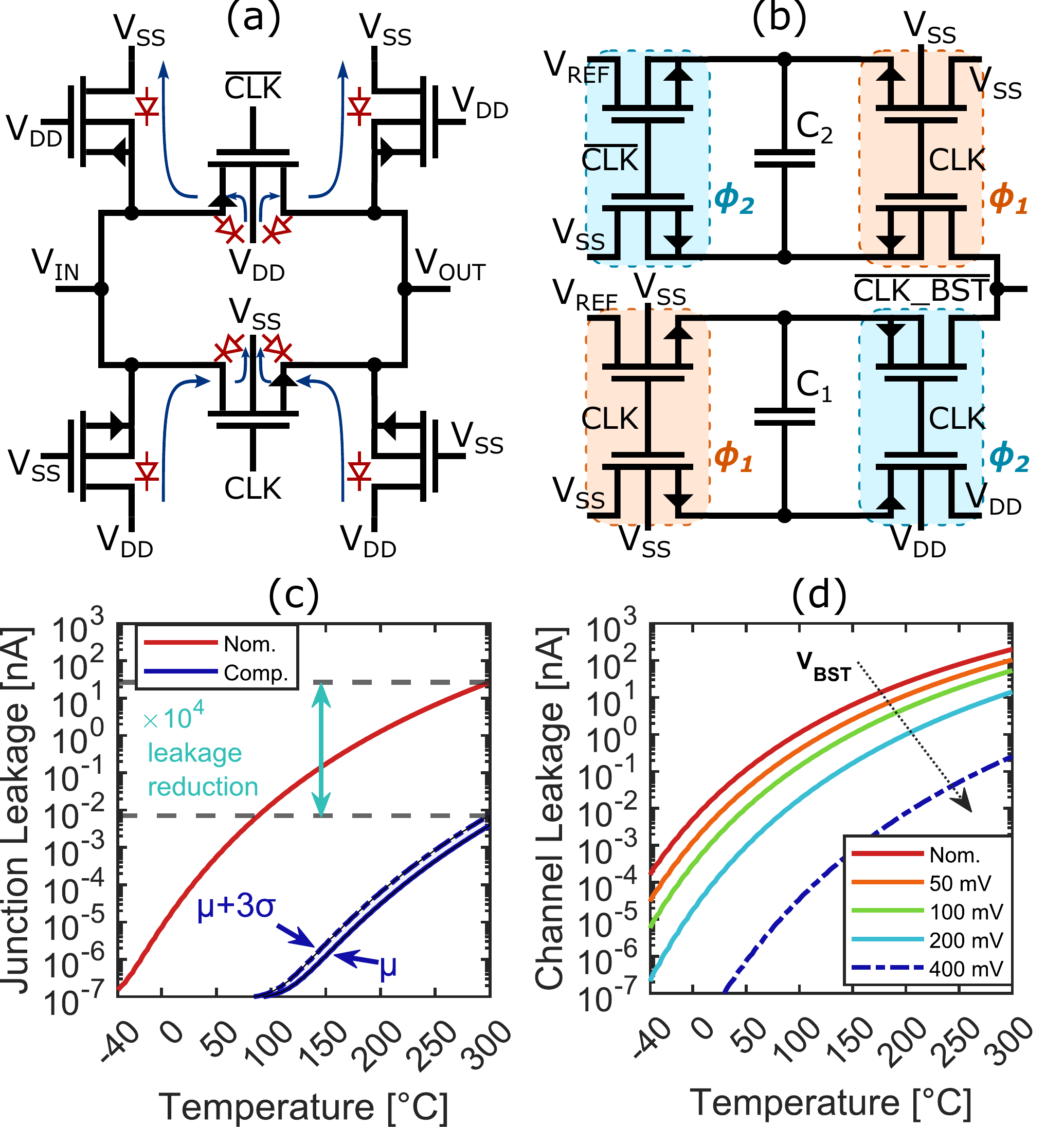}}
\caption{Design techniques for leakage current reduction: (a) Pass-gate circuit with four dummy transistors (on the corners) for junction leakage compensation; (b) Two-phases clock boosting generator circuit; (c) Junction leakage current dependence on temperature and compensation with dummy devices; (d) Channel leakage current dependence on temperature and clock boosting.}
\label{fig_LeakFull}
\end{figure}

The same method can also be applied to the transistors in the central positions of the circuit, for example to the differential input pair. 
However, since these two MOSFETs are generally large, doubling them to realize the compensation yields an increase of the parasitic capacitance on the common node, reducing the speed of the whole circuit.
With the presented design technique, a single dummy device with the same size as one of the transistors in the input pair is enough to provide the necessary compensation current, since both drain-body and source-body junctions are used to collect the compensation leakage.
Then, this current is mirrored with a ratio of four and sunk from the source common connection of the input pair, while the separated drain connections receive double leakage each.
The transistors in the current mirror have minimum dimensions and are compensated for their leakage current, so they realize the compensation mechanism without increasing the parasitic capacitance of the pair common circuit node. 
However, using a current mirror means introducing non-idealities in the mirrored currents, hardening a perfect cancellation of the leakage currents among dummy and active transistors.
This method is anyway fundamental to reduce the current variation in the differential input pair from bias conditions due to the temperature increase, as its contribution is the most significant one given the pair size.

The pass-gates used in the $\Delta$-$\Sigma$ modulator are similarly compensated with dummy transistors, as reported in Fig. \ref{fig_LeakFull}(a).
Both the p-n junctions of a pass-PMOS or a pass-NMOS need to be compensated with leakage currents from the dummy devices, so for each pass-gate there are four dummy transistors, sourcing and sinking compensation currents on the input and output node. 
The better the compensation on the currents, the more stable the voltages across the capacitors can be during the DT circuit holding times.
The effectiveness of the dummy compensation method appears evident from the simulation results reported in Fig. \ref{fig_LeakFull}(c), where we can observe an improvement of more than four decades, also considering mismatch. Let us stress that simulations are strictly reliable only up to the qualification temperature (175~°C) but the trend is clear also at higher temperature.

Pass-gates are heavily affected by the other leakage mechanism in silicon CMOS, which is the subthreshold channel current, showing an exponential dependence on the overdrive voltage \cite{Roy2003,Sbrana2024}.
This effect is strictly related to the degradation of the subthreshold swing as the temperature increases \cite{Colinge2004,Rudenko2002}, reducing the strength of the off-state for a transistor.
A negative off-state gate voltage shift for the n-MOS switches and a positive one for the p-MOS switches can be exploited: in this way, the channel leakage current can be drastically reduced.
Fig. \ref{fig_LeakFull}(b) reports the charge-pump circuit used to generate the overdrive voltages for the clock-boosting solution.
The charge-pump mechanism is exploited to shift both the high and the low levels of the clock signal, obtaining $V\textsubscript{DD}+V\textsubscript{ref}$ for the high level and $-V\textsubscript{ref}$ for the low one employing $C_1$ and $C_2$. 
An additional benefit of this approach is the on-resistance reduction for both the n-MOS and the p-MOS of the pass-gates, thus allowing a smaller aspect ratio of the transistors and consequently a lower reverse current of the p-n junctions. 
The analysis of the improvements introduced with the clock-boosting technique has been performed by means of simulations shown in Fig. \ref{fig_LeakFull}(d), reporting the channel leakage current as a function of the temperature and for different values of the boosting voltage. 
An absolute boost value of 200 mV enables a reduction of the leakage current by a factor larger than 20, still being compliant with the transistor gate voltage maximum rating.
This value has been used in the design for channel leakage compensation, generated with a resistive partition of the supply voltage since high accuracy is not required.

\begin{figure}[t!]
\centerline{\includegraphics[width=0.5\textwidth]{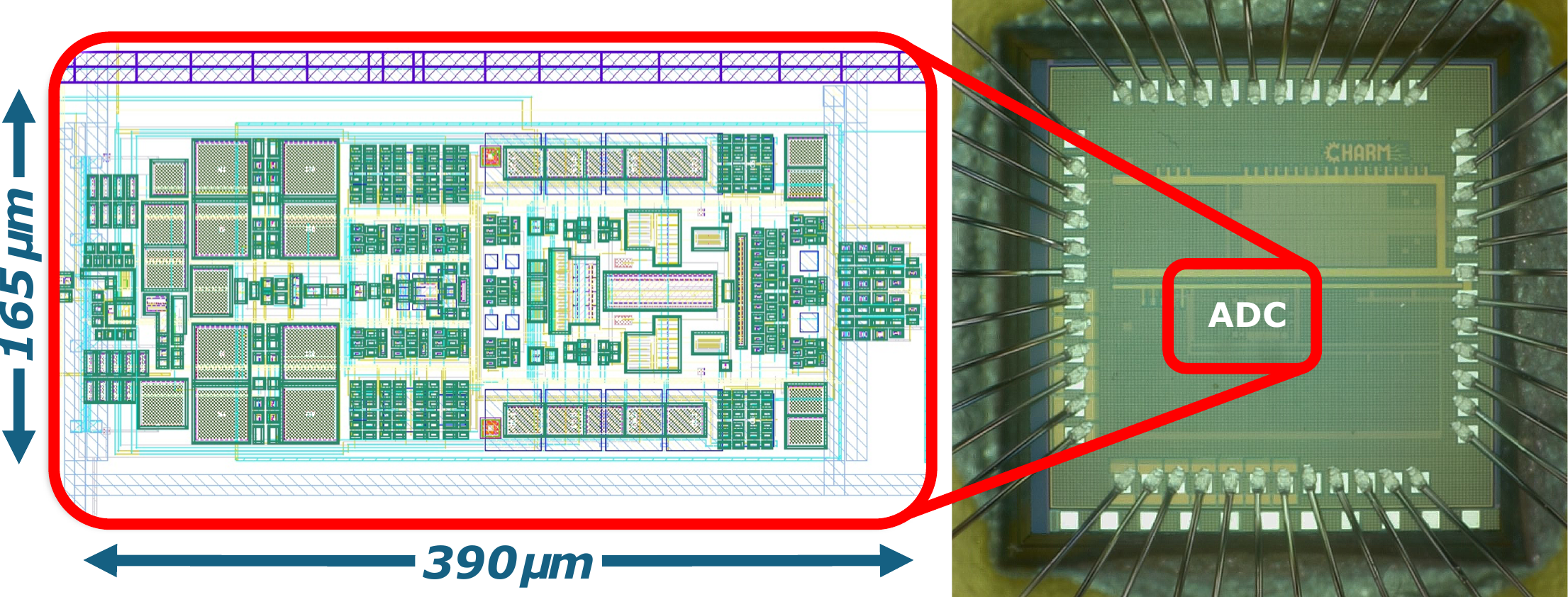}}
\caption{Chip top view and layout magnification on the ADC block.}
\label{fig_ChipLayout}
\end{figure}

The discussed design techniques for both junction and channel leakage compensations have been implemented in this ADC design.
The clock boosting solution has been used to drive all the minimum size pass-gates, while several dummy transistors have been added to many critical nodes of circuits, to either sink or source compensation currents in each switch, bias circuit, op-amp, and comparator. 
Regarding the electromigration problem, we have carefully sized the width of the metal interconnections during the layout phase, so that the current density is a factor ten below the electromigration threshold at 300°C, as extrapolated from available process specifications \cite{Sbrana2024}.
This threshold is about 45~$\mu$A$/\mu$m for the internal metal layers, rising to 75~$\mu$A$/\mu$m for the top metal (with $W = 0.8$~$\mu$m). 

The integration of the presented simple converter architecture with these temperature-aware design techniques results in an on-chip footprint area of 0.065 mm\textsuperscript{2}, and the ADC block is integrated into a more complex IC all optimized for high-temperature operation, as illustrated in Fig.~\ref{fig_ChipLayout}.
The total footprint of the additional circuitry required to enable high-temperature operation is 0.0086~mm\textsuperscript{2}, mostly due to the area of dummy transistors, corresponding to a total area overhead for the extended temperature range of $13.7\%$. 
The increased footprint of the electromigration-resilient metalizations does not lead to a higher area occupation of the ADC.


\section{Chip measurement setup}

\begin{figure}[b!]
\centerline{\includegraphics[width=\columnwidth]{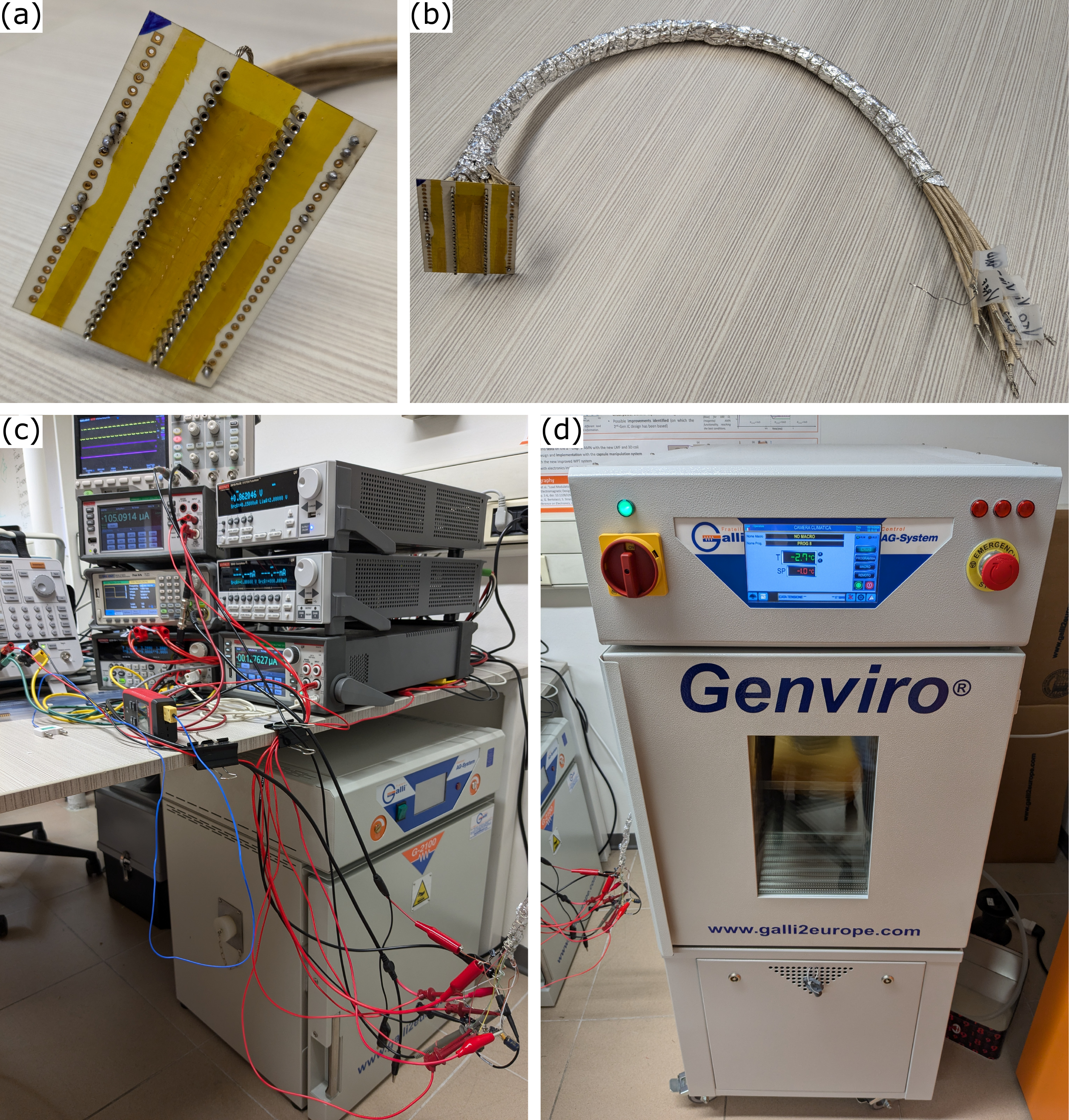}}
\caption{Measurement set-up for ADC characterization in the temperature range \mbox{-40°C$\div$260°C}: (a) Ceramic PCB for chip connection, compatible with the DIL48 package; (b) Connection wires between the ceramic PCB placed inside the oven and the measurement instruments outside it, rated for high temperature and with a custom metal shield for noise reduction; (c) Lab oven for high-temperature measurements up to 260°C, containing the chip with several instruments connected; (d) Climate chamber for low-temperature measurements up to -40°C.}
\label{fig_SetUp}
\end{figure}

The second-order $\Delta$-$\Sigma$ ADC has been designed to operate under the following nominal conditions: clock frequency $f\textsubscript{S}$ of $150~\rm{kHz}$, OSR of $512$, reference voltage $V\textsubscript{ref}$ of $1.8~\rm{V}$ and input common-mode voltage $V\textsubscript{ic}$ of $0.9~\rm{V}$.



The measurement setup used for the temperature characterization of the converter is shown in Fig.~\ref{fig_SetUp}.
The IC has been bonded in a ceramic DIL48 package, able to withstand temperatures as high as 300~°C.
The packaged chip has been placed on a two-face ceramic PCB with a custom metal socket, compliant with the explored temperature range.
The board is then connected to various measurement instruments through wires exploiting high-temperature mica insulation. 
Long wires have been enclosed in a custom metal shield connected to the ground to reduce the electromagnetic noise from the measurement environment.
A laboratory oven and a climatic chamber have been employed to cover the entire temperature range: G2100 and Genviro 030T, respectively \cite{Galli}. G2100 is a 350°C laboratory oven with internal forced ventilation, which includes a lateral opening for routing.
This oven can only heat the internal chamber starting from room temperature, and it has been used to cover the higher part of the considered temperature range.
The other oven is a climate chamber with forced ventilation and a refrigerant circuit, used for low-temperature characterization down to -40°C.
The electrical characterization test setup included the following measurement instruments: a power supply (Keihtley 2230-30-1) to provide the analog core voltage of 1.8 V, I/O voltage of 3.3V and ADC reference voltage; a source measurement unit (Keithley 2601B) to provide the bias current to the chip; two arbitrary waveform generators (Tektronix AFG31102 and Siglent SDG2122X) for clock sources (system, CMFB) and for input AC signals; an oscilloscope (Siglent SDS1204XE) for the converter digital output bitstream acquisition.
Providing all the necessary signals to the ADC from the outside helps to isolate the dependence of its performance on the temperature variations to the contribution introduced by other circuit blocks, such as internal bias current generator, clock generator and voltage reference.


\begin{figure*}[!b]
\centering
\centerline{\includegraphics[width=2.0\columnwidth]{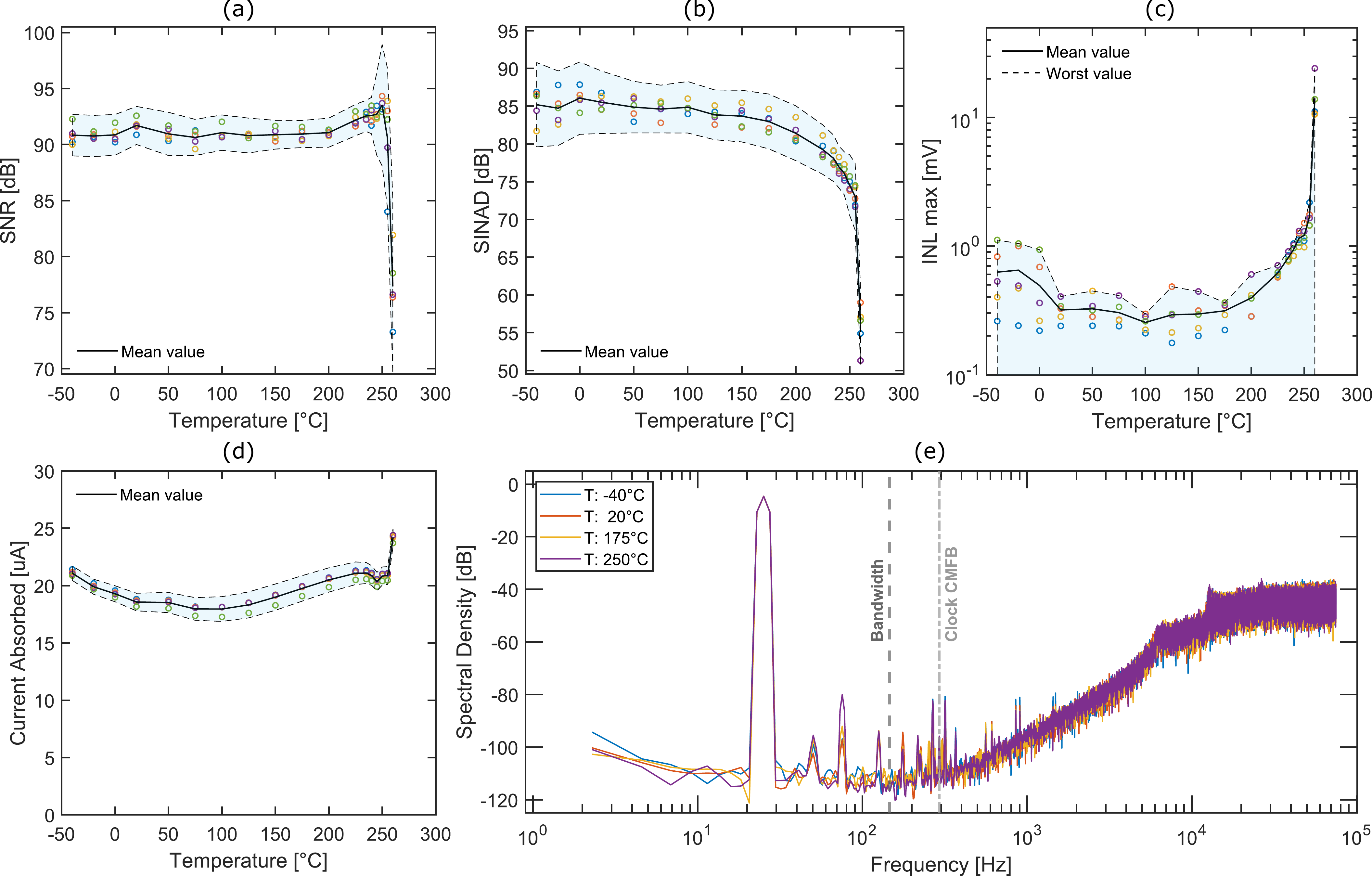}}
\caption{Results from repeated measurements on five different ICs, performed at $f_{S} = 150 kHz$, $OSR = 512$, $f_{CMFB} = f_{S}/512$, $I_{BIAS} = 50 nA$, $V_{REF} = 1.8 V$: (a) Measured SNR and (b) SINAD when an input AC tone is applied ($V_{AC} = 1.7 V$, $f_{AC} = 25.177 Hz$); (c) Worst INL measured from DC characteristics at each temperature step; (d) ADC current consumption on 1.8-V line; (e) Spectrum reported from AC characterization (AC tone at $V_{AC} = 1.7 V$, $f_{AC} = 25.177 Hz$), for a single chip and significant temperature values. The graphs from (a) to (d) report the actual results indicated by circles, the mean trend with a solid line and the $+/-3\sigma$ range delimited by the dashed lines and coloured in light blue.}
\label{fig_meas}
\end{figure*}

Since the XFAB XT018 process is qualified up to 175°C, PDK simulations can not be trusted above this limit, and only accurate high-temperature measurements can confirm the expected behaviour of the circuit.


\section{Measurement results}

The ADC performance has been experimentally assessed through the SINAD, the signal-to-noise ratio (SNR), and the effective number of bits (ENOB) figures of merit measured over temperature for 5 different samples.

The experimental ADC SNR and SINAD as a function of the operating temperature are shown in Fig. \ref{fig_meas} (a) and (b) respectively, indicating the mean with a solid line and the mean plus or minus three standard deviations with dashed lines. 
It can be seen that the SNR is quite stable within the temperature range, reaching a value above 93 dB at 250°C, while it falls down quickly as the operating temperature approaches 260°C. 
A similar trend has been noticed for the SINAD, which however shows a soft worsening from 85 dB at room temperature to 74.5~dB at 250°C, and then collapsing at 260°C.
As the performance degradation in the 100~°C-250~°C range is observed only from the SINAD point of view, but not in the SNR, we can conclude that this is only related to the increasing impact of distortion with rising temperature.
In both charts, the dramatic performance drop as the temperature exceeds 250°C is probably due to a critical temperature-induced alteration of the behavior of some internal circuits and fundamental parts of the ADC architecture.
Even though the main bias signals are provided from the outside, when the temperature increases every p-n junction in the chip is affected by increased junction leakage current, therefore even the pad protection diodes are injecting and withdrawing leakage currents on the input and output connections. 
For this reason, the measured biasing current cannot be trusted at high temperature, and we thus speculate that this could be the main cause of performance degradation. 
In addition, some nominally-off peripheral circuit blocks could turn on unintentionally interfering with signals provided by the characterization setup.

The integrative non-linearity (INL) is obtained from the DC characterization for the considered temperature range. Fig. \ref{fig_meas}(c) shows that the INL increases as the temperature rises, in agreement with the degradation of the AC parameters. 
The measured value is below 1~mV from -40~°C to 250~°C in the worst case among the tested chips, while it deteriorates at temperatures above the upper limit.

Fig. \ref{fig_meas}(d) shows the measured supply current adsorbed by the ADC with $V_{\rm DD} = 1.8$~V for five different samples. As can be seen, the power consumption has a relatively small variation as a function of temperature also above the process qualification temperature and up to 250~°C. 
The effects of both noise and non-linearity can also be seen in Fig. \ref{fig_meas}(e), where the experimental spectral power density is reported for one chip at different temperature values. 
The spectra are quite stable in temperature inside the signal bandwidth region, with low flicker noise. 
When the temperature rises, the distortion peaks become more visible, and the distortion power related to them contributes to reducing the SINAD.
The ground noise is higher than what is expected from simulations, but in high-temperature measurements, no capacitors can be placed close to the chip for an effective filtering action, therefore higher noise is captured from the long and unshielded lines.
The presence of additional spikes can also be noticed in the reported spectra, related to off-band frequencies. 
This additional distortion source is due to the amplifier CMFB control circuit, since it is based on passive switched capacitor networks working at $f_{S}/512$, a frequency that is modulated by the ADC input signal one, leading to two peaks very close to each other.
However, the CMFB frequency is outside the signal bandwidth, and the related distortions can be pushed to higher frequencies by simply increasing the CMFB operating frequency, without loss of performance.

\begin{table}[t!]
\centering
\caption{State-Of-Art High-Temperature ADC Comparison}
\resizebox{\columnwidth}{!}{%
\begin{tabular}{|l|c|c|c|c|c|}
\hline
Reference & {\textbf{This work}} & \cite{Korotkov2020} & \cite{Kappert2022} & \cite{Watson2014} & \cite{Zou2015} \\ \hline
Tech. [$\mu$m] & \textbf{SOI 0.18} & SOI 0.18 & SOI 0.35 & AD7981 & SOI 1.0 \\ \hline
$f\textsubscript{S}$~[kHz] & \textbf{150} & 10000 & 31.25 & 600 & 50 \\ \hline
BW [kHz] & \textbf{0.146} & 100 & 0.488 & 300 & 25 \\ \hline
OSR & \textbf{512} & 50 & 64 & N/A & N/A \\ \hline
SNR @T\textsubscript{MAX} & \textbf{93.4 dB} & - & 84 dB & 88 dB & 56 dB \\ \hline
SINAD @T\textsubscript{MAX} & \textbf{74.5 dB} & 68 dB & 80 dB & 87 dB & 43 dB \\ \hline
ENOB @T\textsubscript{MAX} & \textbf{12 bit} & 11 bit & 13 bit & 14 bit & 6.9 bit \\ \hline
$T\textsubscript{MAX}$ & \textbf{250°C} & 175°C & 250°C & 210°C & 300°C \\ \hline
Power [mW] & \textbf{0.044} & 26.3 & - & 8 & 2.17 \\ \hline
Area [mm\textsuperscript{2}] & \textbf{0.065} & 2.76 & 1.75 & - & 2.5 \\ \hline
Architecture & \textbf{2\textsuperscript{nd} $\Delta$-$\Sigma$} & 2\textsuperscript{nd} $\Delta$-$\Sigma$ & 2\textsuperscript{nd} $\Delta$-$\Sigma$ & SAR & SAR \\ \hline
$V\textsubscript{DD}$ & \textbf{1.8 V} & 3.3 V & 3.3 V & 2.5 V & 5 V \\ \hline
FoM\textsuperscript{(1)} @T\textsubscript{MAX} & \textbf{140} & 134 & N/A & 163 & 114 \\ \hline
\multicolumn{6}{l}{\textsuperscript{(1)}: Schreier FoM defined as SINAD~$+10\times\log(\rm{BW}/P)$.}
\end{tabular}%
}
\label{tabSOA}
\end{table}


The performance extracted from ADC characterization has been reported in Table \ref{tabSOA} for a comparison with other works in the literature.
As already mentioned, the relatively low signal bandwidth has been a design choice to optimize the resolution at the cost of conversion speed, in line with the ADC target application.
The high values of SNR and SINAD at 250~°C, identified as the ADC maximum operating temperature, result in an ENOB of 12 bits.
These results, together with the extremely low area occupation and power consumption, have been used to calculate the Schreier Figure of Merit (FoM) \cite{Schreier2005,Devarajan2009}, reaching a value of 140 dB at 250~°C, and showing the  potential of this high-temperature design. 
Among the data converters for very high temperature, this work performs very well, featuring record Schereier FoM compared to all comparison circuits, except for a commercial SAR converter designed by Analog Devices (AD7981), which however has a much lower maximum temperature of 210~°C, since the used FoM does not consider the temperature range as a key feature.

\section{Conclusions}

We have presented a second-order discrete-time delta-sigma ADC designed for high-temperature operation up to 250~°C with a standard automotive-grade SOI CMOS process qualified up to 175~°C showcasing impressive performance, using a set of design techniques to reduce the negative impact of high temperature on circuit behavior. The temperature-range extension techniques require a total overhead of die area of only 13.7\% for a record-low ADC die area of only 0.065~mm$^2$. The ADC exhibits at 250~°C a SINAD of 74.5 dB, a 12-bit ENOB, and a power consumption of only 44~mW.

The achieved results position this high-temperature design as a leading contender in the field of data converters for extreme environments, with a calculated Schreier Figure of Merit of 140 dB at 250~°C, highlighting its potential for applications in harsh environments. 

This work opens up new possibilities for the design of electronic systems capable of reliable operation in challenging and high-temperature conditions, using standard automotive-grade processes. We do not see fundamental reasons not to further improve these design techniques as to reach the operating temperature of 300~°C with SOI CMOS. Additional gains can be obtained with temperature-resistant interconnect material.

The future adoption of SiC CMOS scaled-down technologies for monolithic integration will enable to reach higher temperature and higher currents. Until such technology becomes available and at a reasonable cost, the capability to design high-temperature circuits in SOI-CMOS is a very cost-effective way to obtain components and systems with small form factors to enable the digitization of several industrial processes and applications.


\bibliographystyle{IEEEtran} 

\begin{IEEEbiography}[{\includegraphics[width=1in,height=1.25in,clip,keepaspectratio]{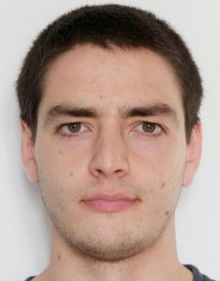}}]{Christian Sbrana}
received the B.S. and M.S. degrees (cum laude) in electronic engineering from the University of Pisa, Italy, in 2018 and 2021, respectively. 
He is currently pursuing the Ph.D. degree in electronics with the University of Pisa, working on the industrial IoT sensors for harsh environments, wireless power transfer systems for implantable medical devices and analog-based physical unclonable functions (PUFs). 
He is also with Quantavis s.r.l. 
His research interests include power management and integrated design on WBG materials, analog and mixed-signal ICs design and sensors for harsh environments.
\end{IEEEbiography}

\begin{IEEEbiography}[{\includegraphics[width=1in,height=1.25in,clip,keepaspectratio]{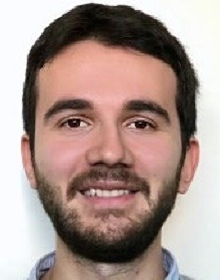}}]{Alessandro Catania}
received the B.S., M.S., and Ph.D. degrees in electronic engineering from the University of Pisa, Italy, in 2014, 2016 and 2020, respectively. 
He is currently working as an Assistant Professor with the University of Pisa. 
His current research interests include mixed-signal microelectronic design for harsh environments and wireless power transfer systems for implantable systems. 
\end{IEEEbiography}

\begin{IEEEbiography}[{\includegraphics[width=1in,height=1.25in,clip,keepaspectratio]{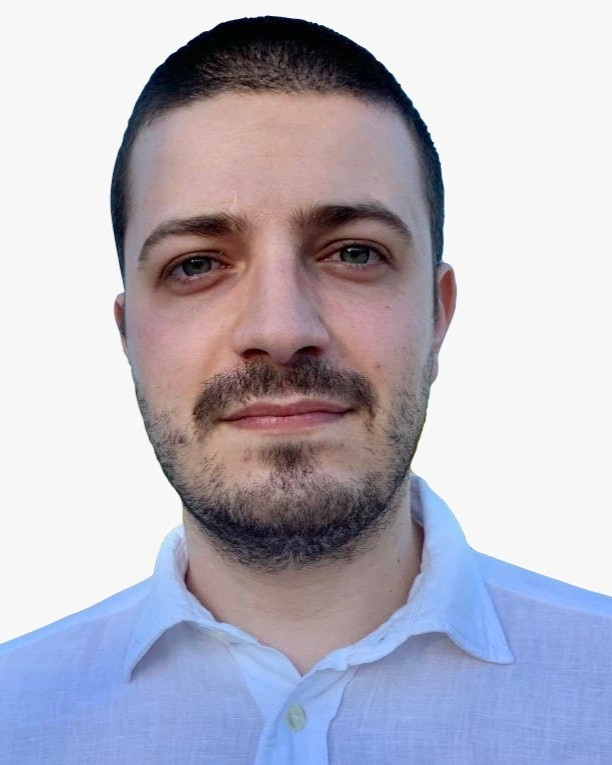}}]{Tommaso Toschi}
received the B.S. and the M.S. degrees in electronic engineering from the University of Pisa, Italy, in 2018 and 2021, respectively.
He has worked for Quantavis s.r.l. designing mixed-signal integrated circuits for harsh environments. 
\end{IEEEbiography}

\begin{IEEEbiography}[{\includegraphics[width=1in,height=1.25in,clip,keepaspectratio]{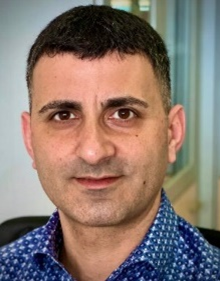}}]{Sebastiano Strangio}
received the B.S. and M.S. degrees (cum laude) in EE, and the Ph.D. degree from the University of Calabria, Cosenza, Italy, in 2010, 2012, and 2016, respectively. 
He was with IMEC, Leuven, Belgium, as a Visiting Student, in 2012, working on the electrical characterization of resistive-RAM memory cells. 
From 2013 to 2016, he was a Temporary Research Associate with the University of Udine, and with the Forschungszentrum Jülich, Germany, as a Visiting Researcher, in 2015, researching on TCAD simulations, design, and characterization of TFET-based circuits. 
From 2016 to 2019, he was with LFoundry, Avezzano, Italy, where he worked as a Research and Development Process Integration and Device/TCAD Engineer, with main focus on the development of a CMOS Image Sensor Technology Platform. 
He is currently a Researcher in electronics with the University of Pisa. 
He has authored or coauthored over 40 papers, most of them published in IEEE journals and conference proceedings. 
His research interests include technologies for innovative devices (e.g. TFETs) and circuits for innovative applications (CMOS analog building blocks for DNNs), as well as CMOS image sensors, power devices and circuits based on wide-bandgap materials. 
\end{IEEEbiography}

\begin{IEEEbiography}[{\includegraphics[width=1in,height=1.25in,clip,keepaspectratio]{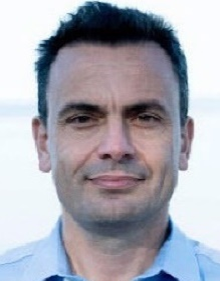}}]{Giuseppe Iannaccone}
received the M.S. and Ph.D. degrees in EE from the University of Pisa, in 1992 and 1996, respectively. 
He is currently Deputy President and Professor of electronics with the University of Pisa. 
He has coordinated several European and national projects involving multiple partners and has acted as principal investigator in several research projects funded by public agencies at the European and national level, and by private organizations. 
He co-founded the academic spinoff Quantavis s.r.l. and is involved in other technology transfer initiatives. 
He has authored or coauthored more than 250 articles published in peer-reviewed journals and more than 160 papers in proceedings of international conferences, gathering more than 11000 citations on the Scopus database. 
His research interests include quantum transport and noise in nanoelectronic and mesoscopic devices, development of device modeling tools, new device concepts and circuits beyond CMOS technology for artificial intelligence, cybersecurity, implantable biomedical sensors, and the Internet of Things. 
He is a fellow of the American Physical Society. 
\end{IEEEbiography}


\end{document}